\documentclass{ws-ijmpd0}
\usepackage{bm}
\usepackage{amssymb, amsfonts, amsbsy}
\def\TODAY{6 May 2010; \LaTeX-ed \today}

\begin{document}

\markboth{Jozef Sk\'akala and Matt Visser}
{Pseudo--Finslerian spacetimes and multi-refringence}

%
\catchline{}{}{}{}{}
%
\title
{Pseudo--Finslerian spacetimes and multi-refringence}
\author{Jozef Sk\'akala and Matt Visser}
\address{School of Mathematics, Statistics, and Operations Research, 
Victoria University of Wellington, Wellington, New Zealand\\ {\sf \{jozef.skakala, matt.visser\}@msor.vuw.ac.nz}  }
\maketitle
\begin{abstract}
Ongoing searches for a ``quantum theory of gravity'' have repeatedly led to the suggestion that space-time might ultimately be anisotropic (Finsler-like) and/or exhibit ``multi-refringence'' (multiple ``signal cones'').  While multiple (and even anisotropic) signal cones can easily be dealt with in a unified manner, by writing down a \emph{single} ``Fresnel equation'' to  simultaneously encode all signal cones in an even-handed manner, once one moves ``off the signal cone'' and attempts to construct a full multi-refringent ``spacetime metric'' we shall see that the situation becomes considerably more problematic.  (In contrast, in general relativity, the single [mono-refringent] signal cone, together with a single scalar degree of freedom, the conformal factor, uniquely characterizes the full Lorentzian spacetime metric.)   In fact in the multi-refringent case we shall report a significant \emph{negative} result, a ``no-go'' argument: Despite the fact that multiple signal cones can easily be unified in a Fresnel formalism,  multi-refringent spacetime metrics strongly resist any attempt at unification --- exhibiting serious and wide-spread pathologies that appear intrinsic, comprehensive,  and unavoidable.

To throw some light on this situation we shall work by analogy --- we shall  explore the one physical situation where we have direct experimental evidence that both of these phenomena (anisotropy and multi-refingence) occur simultaneously: \emph{bi-axial bi-refringent crystal optics}. We shall first develop a \emph{spacetime} interpretation of the standard purely spatial Finsler structures, leading to the notion of pseudo--Finsler spacetimes. 
We shall then study the extent to which we can ``unify'' the spacetime treatment of multiple polarization modes by adopting  a Lorentzian-signature generalization of Riemann's prescription, and attempting to encode everything in one overarching pseudo-Finsler \emph{spacetime metric}.  We find that the natural ``unified'' quantity, while it defines a perfectly good pseudo-Finsler \emph{norm}, leads to a seriously diseased pseudo-Finsler \emph{metric}. From the pseudo-Finsler norm we can easily set up a Fresnel equation, but the attempt to go ``off-shell'' to construct a full spacetime pseudo-Finsler geometry is doomed.

The significance of this result extends beyond the optical framework in which (purely for pedagogical and illustrative  reasons) we are working,  and has implications for a wide class of models that attempt to introduce multi-refringence and Finsler-like intrinsic anisotropies to the low-energy manifold-like limit of  ``quantum gravity''.
Our ``no-go'' result indicates that in multi-refringent models  there is no simple or compelling way to construct any unifying notion of pseudo--Finsler spacetime metric. At the level of the metric (though not at the level of the signal cones) one has to resort to multiple independent pseudo-Finslerian metrics, often with troublesome ``cusps'' on the individual signal cones. 
This strongly suggests that, despite its initial popularity, multi-refringence is not a useful route to explore.


\end{abstract}

\def\d{{\mathrm{d}}}
\def\implies{\Rightarrow}
\newcommand{\scri}{\mathscr{I}}
\newcommand{\sun}{\ensuremath{\odot}}
\def\ep{\epsilon}
\def\k{\mathbf{k}}
\def\x{\mathbf{x}}
\def\v{\mathbf{v}}
\def\s{\mathbf{s}}
\def\e{\mathbf{e}}
\def\t{\mathbf{t}}
\def\n{\mathbf{n}}
\def\u{\mathbf{u}}
\def\w{\mathbf{w}}
\def\eg{{\it e.g.}}
\def\ie{{\it i.e.}}
\def\etc{{\it etc.}}
\def\sign{{\hbox{sign}}}
\def\eof{\Box}
\newenvironment{warning}{{\noindent\bf Warning: }}{\hfill $\eof$\break}

\clearpage
\section{ Introduction}

The theoretical physics community has recently exhibited increasing interest in the possibility of ultra-high-energy violations of Lorentz invariance,\cite{Jacobson}\cdash\cite{SVW} and in the possibility of quantum-gravity-induced spacetime anisotropies and multi-refringence. 
Specifically --- recent speculations regarding Lorentz symmetry breaking and/or fundamental anisotropies and/or multi-refringence arise separately in the many and various approaches to quantum gravity. Such phenomena arise in loop quantum gravity,\cite{LQG} string models,\cite{string} and causal dynamical triangulations,\cite{causal} and are also part and parcel of the ``analogue spacetime'' programme,\cite{analogue} and of many attempts at developing ``emergent gravity''.\cite{Kostelecky}\cdash\cite{emergence}  Recently, the ultra-high energy breaking of Lorentz invariance has been central to the Horava--Lifshitz models.\cite{Horava}\cdash\cite{SVW} Of course not all models of quantum gravity lead to high-energy Lorentz symmetry breaking, and the comments below should be viewed as exploring one particular class of interesting models. 

If we wish to develop a geometric spacetime framework for representing Lorentz symmetry
breaking, either due to spacetime anisotropies or multi-refringence, then it certainly cannot be standard pseudo--Riemannian geometry. This strongly suggests that  carefully thought out extensions and modifications of pseudo--Riemannian geometry might be of real interest to  both the general relativity and high-energy communities. 
In particular, when attempting to generalize pseudo-Riemannian geometry the interplay between the ``signal cones''  of a multi-refringent theory and the generalized spacetime geometry is an issue of considerable interest:
\begin{itemize}
\item 
In standard general relativity the (single, unique) signal cone almost completely specifies the spacetime geometry --- one need only supplement the signal cone structure with one extra degree of freedom at each point in spacetime,  an overall conformal factor,  in order to completely specify the spacetime metric, and thereby completely specify the geometry. (This is ultimately due to the fact that in standard pseudo-Riemannian geometry the scalar product is a simple bilinear operation.) 

\item
In multi-refringent situations it is quite easy to unify all the signal cones in one single Fresnel equation that simultaneously describes all polarization modes on an equal footing.

\item 
In a standard manifold setting, where we retain the usual commutative coordinates, we shall see that it is natural to demand that each polarization mode can be assigned an independent pseudo-Finsler metric. Earlier work on Finsler-like applications to physics includes that of Takano\cite{Takano}, Asanov\cite{Asanov}, Vacaru\cite{Vacaru1}\cdash\cite{Vacaru2},  and others.\cite{Brandt}\cdash\cite{Gonner} More recent work has appeared in references~\refcite{Vacaru3}--\refcite{Chang}, and other articles cited below.

\item
What is exceedingly difficult, and we shall argue is in fact outright impossible within this framework, is to  construct a \emph{unified} formalism that moves ``off-shell'' (off the signal cones). 

\item 
This is a \emph{negative} result, a ``no-go theorem'',  which we hope will focus attention on what can and cannot be accomplished in any natural way when dealing with multi-refringent anisotropic Finsler-like extensions to general relativistic spacetime.

\item 
If one steps outside of the usual manifold picture, either by adopting non-commutative coordinates, or even more abstract choices such as spin foams, causal dynamical triangulations, or string-inspired models, then the issues addressed in this article are moot --- our considerations are relevant only insofar as one is interested in the first nontrivial deviations from exact low-energy Lorentz invariance, and only relevant insofar as these first nontrivial effects can be placed in a Finsler-like setting.

\end{itemize}
Considerable insight into such Finsler-like models can be provided by considering the ``analogue spacetime'' programme, where analogue models of curved spacetime emerge at some level from well understood physical systems.\cite{analogue}
In particular, the physics of \emph{bi-axial bi-refringent crystals}\cite{BW} provides a particularly simple physical analogue model for the mathematical object introduced some 155 years ago by Bernhard Riemann,\cite{Riemann}  and now known as a Finsler distance. (See \ref{A:Riemann} for some general mathematical background regarding Finsler spaces.) 

We shall soon see that this mathematical object can reasonably easily be extended to a Lorentzian-signature pseudo-Finsler space-time, with an appropriate pseudo-Finsler norm, though defining a pseudo-Finsler metric is more subtle and fraught with technical problems. 
(It must be emphasized that,  despite many misapprehensions to the contrary,  \emph{uni-axial} birefringent crystals are relatively uninteresting in this regard; they do not lead to Finsler 3-spaces, but ``merely'' yield bi-metric Riemannian 3-geometries.)  

While the use of Finsler 3-spaces to describe crystal optics is reasonably common knowledge within the community of mathematicians and physicists studying Finsler spaces, it is very difficult to get a clear and concise explanation of exactly what is going on when one generalizes to Lorentzian signature space-time. In particular the fact that any relativistic formulation of Finsler space needs to work in Lorentzian signature $(-+++)$, instead of the Euclidean signature $(++++)$ more typically used by the mathematical community, leads to many technical subtleties (and can sometimes completely invalidate naive conclusions). 

Note that for the purposes of the current article we are not particularly interested in the properties of birefringent crystals \emph{per se}, we are really only interested in them as an \emph{exemplar} of Finsler 3-space and Finsler space-time, as a guidepost to more complicated things that may happen in Finslerian extensions to general relativity. We shall already see a number of interesting pitfalls arising at this (in principle) quite elementary level.

Our ultimate goal is to be able to say something about the (presumed) low-energy manifold-like limit of whatever quantum theory, or class of quantum theories,  approximately reproduces Einstein gravity. To this end, our ``no-go'' result indicates that the popular assumption that anisotropies and multi-refingence are likely to occur in ``quantum gravity'' leads to significant difficulties for the Einstein equivalence principle --- since even the loosest interpretation of the Einstein equivalence principle would imply the necessity of a coherent formalism for dealing with all signal comes, and the spacetime geometry, in some unified manner. 

We conclude that, despite the fact that spacetime anisotropies and multi-refringence are very popularly assumed to be natural features of ``quantum gravity'', and while these features have a  straightforward  ``on-shell'' implementation in terms of a suitably defined Fresnel equation, there is no natural way of extending them ``off-shell'' and embedding them into a single over-arching spacetime geometry.

\section{Outline}

Purely for the purposes of developing a useful analogy, which we shall use as a guide to the mathematics we wish to develop, we will focus on the optical physics of \emph{bi-axial bi-refringent crystals}.\cite{BW}  After very briefly presenting the basic definitions, we show how various purely spatial \emph{3-space} Finsler structures arise. (Many purely technical details, when not directly  involved in the logic flow,  will be relegated to the appendices.) 
We again emphasize that \emph{uni-axial} bi-refringent crystals, which are what much of the technical literature and textbook presentations typically focus on, are for our purposes rather uninteresting --- uni-axial bi-refringent crystals ``merely'' lead to bi-metric Riemannian space-times and are from a Finslerian perspective ``trivial''. 

We shall soon see that even in three-dimensional \emph{space} there are \emph{at least four} logically distinct Finsler structures of interest:  On the tangent space each of the two photon polarizations leads, via study of the \emph{group velocity}, to two quite distinct Finsler spacetimes.  On the co-tangent space each of the two photon polarizations leads, via study of the \emph{phase velocity}, to two quite distinct co-Finsler spacetimes. The inter-relations between these four structures is considerably more subtle than one might naively expect. 

Additionally, (apart from some purely technical difficulties along the optical axes in bi-axial crystals), each of these four 3-dimensional spatial Finsler structures has a natural 4-dimensional extension to a \emph{spacetime} pseudo-Finsler structure. Beyond that, there are reasonably natural ways of merging the two photon polarizations into  ``unified'' Finsler and co-Finsler \emph{norms}, closely related to the appropriate Fresnel equation, though the associated Finsler \emph{metrics} are considerably more problematic --- all these mathematical constructions do come with a price --- and we shall be careful to point out exactly where the the technical difficulties lie.

Finally, using this well-understood physical system as a template, we shall (in the spirit of analogue spacetime programme\cite{analogue}) then ask what this might tell us about possible Finslerian extensions to general relativity, and in particular to the subtle relationship between bi-refringence and bi-metricity,\cite{birefringence}$^,$\cite{lnp} (or more generally, multi-refringence and multi-metricity).

Specifically, we have investigated the possibility of whether it is possible to usefully and cleanly deal with both Finsler structure (anisotropy) and multi-refringence \emph{simultaneously}. That is, given two (or more) ``signal cones'': Is it possible to naturally and intuitively construct a  ``unified'' pseudo-Finsler spacetime such that the pseudo-Finsler \emph{metric} specifies null vectors on these ``signal cones'', \emph{but has no other zeros or singularities}? Our results are much less encouraging than we had originally hoped, for and lead to a ``no-go'' result. Our analysis suggests suggest that while pseudo-Finsler spacetimes are certainly useful constructs, they do not appear suitable for dealing with multi-refringence --- when trying to analyze the full spacetime geometry (instead of merely restricting attention to the signal cones) it seems physically more appropriate to think of physics as taking place in a single topological manifold that carries several distinct pseudo-Finsler metrics, one for each polarization mode.

\section{Finsler basics}

Mathematically, a Finsler function (Finsler norm, Finsler distance function) is defined as a function $F(x,v)$ on the tangent bundle to a manifold such that,\cite{Finsler}\cdash\cite{Bejancu} 
\begin{equation}
F(x, \kappa \, v) = \kappa \; F(x, v).
\end{equation}
This then allows one to define a notion of distance on the manifold, in the sense that
\begin{equation}
S\left(x(t_i),x(t_f)\right) = \int_{t_i}^{t_f}  F\left( x(t),  {\d x(t) \over\d t} \right) \; \d t
\end{equation}
is now guaranteed to be independent of the specific parameterization $t$. In the case of a (pseudo--)Riemannian manifold with metric $g_{ab}(x)$ one would take
\begin{equation}
F(x, v) = \sqrt{ g_{ab}(x)\; v^a\,v^b},
\end{equation}
but a general (pseudo--)Finslerian manifold the function $F(x, v)$ is arbitrary except for the linearity constraint in $v$. Note that in Euclidean signature, the general Finsler function $F(x,v)$ is typically smooth except at $v=0$. In Lorentzian signature however, $F(x,v)$ is typically non-smooth for all null vectors --- so that non-smoothness issues have grown to affect (and infect) the entire null cone (signal cone). As we shall subsequently see below, sometimes a suitable higher algebraic \emph{power}, $F^{2n}(x,v)$,  of the pseudo-Finsler norm is smooth. 

In significant contrast to the above, to ensure smoothness of the Finsler \emph{metric}, defined below, it is advisable (even in the Euclidean-signature case, let alone in Lorentzian signature) to go beyond the smoothness of $F(x,v)$ and to demand the considerably stronger condition that the square $F^2(x,v)$ be smooth, except possibly at $v=0$.
It is standard to define the (pseudo--)Finsler \emph{metric} as
\begin{equation}
g_{ab}(x,v) = {1\over2} \; { \partial^2 [F^2(x,v)] \over \partial v^a \; \partial v^b}
\end{equation}
which then satisfies the constraint
\begin{equation}
g_{ab}(x, \kappa\, v ) = g_{ab}(x, v).
\end{equation}
This can be viewed as a ``direction-dependent metric'', and is clearly a significant generalization of the usual (pseudo--)Riemannian case. Almost all of the relevant mathematical literature has been developed for the Euclidean signature case (where $g_{ab}(x,v)$ is taken to be a positive definite matrix). Because of this assumption, any mathematical result  that depends critically on the assumed positive definite nature of the matrix of metric coefficients \emph{cannot} be carried over into the physically interesting pseudo-Finsler regime, at least not without an independent proof that avoids the positive definite assumption. (Unfortunately it is not uncommon to find significant mathematical errors in the pseudo-Finsler physics literature due to neglect of this elementary point.)   Basic references within the mathematical literature include \refcite{Finsler}--\refcite{Bejancu}.  

There have been several recent attempts at using Finsler geometry in various aspects of current physics
research. For example, one use is related to anisotropic geometries,\cite{Siparov} another is related to modifications of general relativity and cosmology.\cite{chang-li} In more fundamental work, Finsler geometries have been directly related to Lorentz symmetry breaking in so-called ``very special relativity''.\cite{Gibbons,Kouretsis}  More recently, it has been noted that the geometry of the spatial slices of arbitrary stationary space-times can cleanly be interpreted in terms of the particular class of Finsler geometries known as Randers spaces.\cite{Gibbons2} (One might also try to generalize the spacetime in a slightly different direction by adopting the notion of an ``area metric'',\cite{area} but we shall not take that particular route for now.) 

An application more directly related to our specific topic of interest was made by Duval,\cite{Duval} where he uses 3-dimensional purely \emph{spatial} Finsler geometries (effectively choosing a preferred inertial frame). The use of Finsler geometry for birefringent crystal optics was also analysed from a more mathematical point of view  in reference \refcite{Antonelli3}, but this is again restricted to 3-dimensional,  preferred-frame, purely \emph{spatial}  Finsler geometry. In contrast, we are for current purposes interested in building up a full \emph{(3+1)-dimensional pseudo-Finsler space-time geometry} based on birefringent crystals, as at a minimum such a space-time formulation will certainly be necessary for developing the possible extensions of general relativity we are interested in.

\section{Space versus space-time: Interpretations of the Finsler structure}

The key physics point in bi-axial bi-refringent crystal optics is that the group velocities, and the phase velocities, are both anisotropic, and depend on direction in a rather complicated way.  Technical details that would detract from the flow of the article are relegated to \ref{A:BW}.

\subsection{From group velocity to Finsler structure}

We can summarize the situation by pointing out that the group velocity is given by
\begin{equation}
v_g^2(\n) = {\bar q_2(\n,\n) \pm \sqrt{ \bar q_2(\n,\n)^2 - \bar q_0(\n,\n)\;(\n \cdot \n) }\over \bar q_0(\n,\n)},
\end{equation}
where $\bar q_2(\n,\n)$ and $\bar q_0(\n,\n)$ are known quadratic functions of the direction $\n$. The coefficients in these quadratic forms are explicit functions of the components of the $3\times3$ permittivity tensor.  See equations~(\ref{E:bar-q0}) and (\ref{E:bar-q2}). 
The function $v_g(\n)$ so defined is homogeneous of degree zero in the components of $\n$:
\begin{equation}
v_g( \kappa\, \n ) = v_g(\n) = v_g(\hat\n).
\end{equation}
The homogeneous degree zero property should remind one of the relevant feature exhibited by the Finsler metric. It is also useful to calculate the so-called ``group slowness'':
\begin{equation}
{1\over v_g(\n)^2} =  {\bar q_2(\n,\n) \mp \sqrt{ \bar q_2(\n,\n)^2 - \bar q_0(\n,\n)\;(\n\cdot \n) }\over (\n\cdot\n)}.
\end{equation}
Let us now first define the quantities
\begin{eqnarray}
F_{3\pm}(\n) &=&  {||\n||\over v_g(\n)} 
= \sqrt{\bar q_2(\n,\n) \mp \sqrt{ \bar q_2(\n,\n)^2 - \bar q_0(\n,\n)\;(\n \cdot \n) }},
\end{eqnarray}
or adopt the perhaps more transparent notation
\begin{eqnarray}
F_{3\pm}(\d\x) =  {||\d\x||\over v_g(\d\x)} 
=
\sqrt{\bar q_2(\d\x,\d\x) \mp \sqrt{ \bar q_2(\d\x,\d\x)^2 - \bar q_0(\d\x,\d\x)\;(\d\x \cdot \d\x) }}.
\end{eqnarray}
This is by inspection a 3-dimensional (Riemannian) Finsler distance defined on \emph{space}, having all the correct homogeneity properties, $F_{3\pm}(\kappa\,\d\x) = \kappa\, F_{3\pm}(\d\x)$. Physically, the Finsler distance is in this situation the time taken for the wavepacket to travel a distance $\d\x$. 

One sometimes encounters comments in the literature to the effect that this specific Finsler distance is an example of a Kropina metric. Such comments are unfortunately \emph{false}. If we define $\alpha = \sqrt{g_{ij}\,\d x^i\,\d x^j}$, and $\beta = b_i \,\d x^i$, then Kropina metrics are of the form $F(\d\x)=\alpha^2/\beta$, and the more general class of so-called $(\alpha,\beta)$ metrics is of the form $F(\d\x)=f(\alpha,\beta)$. By inspection, the Finsler geometry arising in bi-axial optics is neither a Kropina metric nor even an $(\alpha,\beta)$ metric. 

In contrast, the Finsler 3-norm $F_{3\pm}$ defined above is certainly compatible with equation~(3.14) of reference \refcite{Duval}. To see this, it is a brief calculation to verify that
\begin{eqnarray}
\label{E:duval}
|| \bar \e_1\times \d\x|| \; ||\bar \e_2\times\d\x|| 
&&\quad
\leftrightarrow  \quad
\sqrt{ \bar q_2(\d\x,\d\x)^2 - \bar q_0(\d\x,\d\x)\;(\d\x \cdot \d\x)},
\qquad
\end{eqnarray}
where $\bar \e_{1,2}$ are the ray optical axes as defined in \ref{A:axes}.  See also equation~(\ref{E:discriminant}) below. Unfortunately in the bi-axial case equation~(3.13) of reference~\refcite{Duval} is not correct, unless you reinterpret the denominator in that equation not (as claimed) as a Riemannian metric, but instead as a Finsler metric in its own right. Nor in the uni-axial case is equation~(3.13) of reference~\refcite{Duval} correct; the correct statement is that in the uni-axial case both ordinary and extraordinary rays are governed by (distinct) but quite ordinary Riemannian geometries.

\bigskip

To now extend the construction given above to full (3+1) dimensional \emph{spacetime}, we first define a generic 4-vector
\begin{equation}
\d X = (\d t; \d\x),
\end{equation}
and then formally construct
\begin{equation}
F_{4\pm}(\d X) = \sqrt{ - (\d t)^2 + F_{3\pm}(\d\x)^2 }.
\end{equation}
That is
\begin{equation}
F_{4\pm}(\d X) 
=
\sqrt{ - (\d t)^2 + {\d\x\cdot\d\x\over v_g(\d\x)^2} }.
\end{equation}
Even more explicitly, one may write
\begin{eqnarray}
F_{4\pm}(\d X)  &=& \Big[ - (\d t)^2 +\bar q_2(\d\x,\d\x)
\mp
\sqrt{ \bar q_2(\d\x,\d\x)^2 - \bar q_0(\d\x,\d\x)\;(\d\x\cdot \d\x) }\Big]^{1/2}.
\qquad
\end{eqnarray}
The null cones (signal cones) of $F_{4\pm}(\d X)$ are defined by
\begin{equation}
F_{4\pm}(\d X) = 0 \qquad \Leftrightarrow \qquad ||\d\x|| = v_g(\d\x) \; \d t.
\end{equation}
So far this has given us a very natural \emph{pair} of (3+1)-dimensional pseudo--Finsler structures in terms of the ray velocities corresponding to the two photon polarizations.  

\bigskip

\noindent
For future use, let us now formally define the quantity
\begin{eqnarray}
\d s^4 &=& \left\{ F_4(\d X) \right\}^4 
\\
&=& \left\{ F_{4+}(\d X) \; F_{4-}(\d X) \right\}^2
\nonumber
\label{E:4d-dc-F}
\\
&=&
(\d t)^4 - 2 (\d t)^2 \; \bar q_2(\d\x,\d\x) + \bar q_0(\d\x,\d\x)\;(\d\x\cdot \d\x).
\nonumber
\end{eqnarray}
This is tantalizingly close to the quartic extension of the notion of distance that Bernhard Riemann speculated about in his inaugural lecture.\cite{Riemann} See \ref{A:Riemann}, and especially equation~(\ref{E:Riemann4}). This is certainly an example of a specific and simple  $4^{th}$-root pseudo-Finsler norm that can naturally and symmetrically be constructed from the two polarization modes, and its properties (and defects) are certainly worth investigationg. Physically the condition $\d s=0$ defines a double-sheeted conoid (a double-sheeted topological cone)   that is the union of the propagation cone of the individual photon polarizations. However we shall soon see that even though this construction is formally very close to Riemann's original suggestion, when it comes to defining a Finsler spacetime \emph{metric} this construction nevertheless leads to a number of severe technical difficulties; difficulties that can be tracked back to the fact that we are working in Lorentzian signature.

\subsection{From phase velocity to co-Finsler structure}
In counterpoint, as a function of wave-vector the phase velocity is
\begin{equation}
v_p^2(\k) = {q_2(\k,\k) \pm\sqrt{ q_2(\k,\k)^2 - q_0(\k,\k) \; (\k\cdot \k) }\over (\k\cdot\k)},
\end{equation}
where the quadratics $q_2(\k,\k)$ and $q_0(\k,\k)$ are now given by equations~(\ref{E:q2}) and (\ref{E:q0}). 
This expression is homogeneous of order zero in $\k$, so that
 \begin{equation}
v_p(\kappa  \, \k) = v_p(\k) = v_p(\hat \k).
\end{equation}
Again, we begin to see a hint of Finsler structure emerging.
Because $\k$ is a wave-vector it transforms in the same way as the gradient of the phase; thus $\k$ is most naturally thought of as living in the 3-dimensional space of co-tangents to physical 3-space. Let us now define a \emph{co-Finsler} structure on that co-tangent space by
\begin{eqnarray}
G_{3\pm}(\k) &=&  v_p(\k)\;  ||\k|| 
=
\sqrt{ q_2(\k) \pm\sqrt{ q_2(\k,\k)^2 - q_0(\k,\k) \; (\k\cdot \k) } }.
\end{eqnarray}
We use the symbol $G$ rather than $F$ to emphasize that this is a co-Finsler structure, and note that this object satisfies the required homogeneity property
\begin{equation}
G_{3\pm}(\kappa \k) =  \kappa\,G_{3\pm}(\k).
\end{equation}
Now let us go for a (3+1) dimensional spacetime interpretation: Consider the 4-co-vector
\begin{equation}
k = \left(\omega; \k \right),
\end{equation}
and define
\begin{equation}
G_{4\pm}(k) = \sqrt{ - \omega^2 + G_{3\pm}(\k)^2 }.
\end{equation}
That is
\begin{equation}
G_{4\pm}(k) =  \sqrt{ - \omega^2 + v_p(\k)^2\; (\k \cdot \k) }.
\end{equation}
More explicitly
\begin{eqnarray}
G_{4\pm}(k)  &=&
 \Big[ - \omega^2 +    q_2(\k,\k) 
 \pm \sqrt{q_2(\k,\k)^2 - q_0(\k,\k) \; (\k\cdot \k)}\Big]^{1/2} .
\end{eqnarray}
We again see that this object satisfies the required homogeneity property
\begin{equation}
G_{4\pm}(\kappa k) =  \kappa\,G_{4\pm}(k),
\end{equation}
so that this object is indeed suitable for interpretation as a co-Finsler structure. Furthermore the \emph{null co-vectors} of $G_4$ are defined by
\begin{equation}
G_{4\pm}(k) = 0 \qquad \Leftrightarrow \qquad \omega = v_p(\k) \; ||\k||,
\end{equation}
which is exactly the notion of \emph{dispersion relation} for allowed ``on mass shell'' wave-4-vectors that we are trying to capture. Thus $G_{4\pm}$ lives naturally on the co-tangent space to physical spacetime, and we can interpret it as a pseudo-co-Finsler structure. 

We can again define a ``unified'' quantity
\begin{eqnarray}
G_{4}(k)^4 &=& \left\{ G_{4+}(k) \; G_{4-}(k) \right\}^2
\nonumber\\
&=&  \omega^4 - 2 \omega^2  \; q_2(\k,\k) + q_0(\k,\k) \; (\k\cdot \k).\quad
\label{E:4d-dc-G}
\end{eqnarray}
Physically, the condition $G_4(k)=0$ simultaneously encodes both dispersion relations for the two photon polarizations. Furthermore, we see that $G_4(k)$ is tantalizingly close to a co-tangent space version of the quartic expression considered by Bernhard Riemann.\cite{Riemann} Again, see \ref{A:Riemann}, and especially equation~(\ref{E:Riemann4}). Physically the condition $G_{4}(k)=0$ defines a double-sheeted conoid (a double-sheeted topological cone) that is the union of the dispersion relations of the individual photon polarizations. The vanishing of $G_{4}(k)$ can be viewed as a Fresnel equation, and can indeed be directly related to Fresnel's condition for the propagation of a mode of 4-wavenumber $k = (\omega; \k)$. 
As is the case for $F_4(\d X)$,  we shall soon see that even though this construction for $G_{4}(k)$ is formally very close to Riemann's original suggestion, this construction (once one tries to extract a spacetime co-Finsler \emph{metric}) nevertheless leads to a number of severe technical difficulties; difficulties that can again be tracked back to the fact that we are now working in Lorentzian signature.

\section{Technical issues and problems}

The situation as presented so far looks very pleasant and completely under control --- and if what we had seen so far were all there was to the matter, then the study of pseudo-Finsler space-times would be very straightforward indeed --- but now let us indicate where potential problems are hiding.
\begin{itemize}
\item
Note that up to this stage we have not established any direct connection between the Finsler functions $F_\pm(\n)$ and the co-Finsler functions $G_\pm(\k)$. Physically it is clear that they must be very closely related, but (as we shall soon see) establishing the precise connection is tricky.

\item 
Furthermore, the transition from Finsler \emph{distance} to Finsler \emph{metric} requires at least two derivatives.  Even in Riemannian signature this places some smoothness constraints on the Finsler distance, smoothness constraints that are nontrivial and not always satisfied. (In Riemannian signature potential problems are confined to the immediate vicinity of the zero vector.) 

\item There are also additional problematic technical issues involving the 4-dimensional \emph{spacetime} pseudo-Finsler and pseudo-co-Finsler metrics --- again certain components of the metric are infinite, but this time the potential pathology is more widespread. (In Lorentzian signature potential problems tend to infect the entire null cone.) 

\end{itemize}

\subsection{Defining Finsler and co-Finsler 3-metrics}

The standard definition used to generate a Finsler \emph{metric} from a Finsler \emph{distance} is to set:
\begin{equation}
g_{ij}(\n) =  {1\over2} {\partial^2 [F_{3\pm}(\n)^2]\over \partial n^i \, \partial n^j},
\end{equation}
which in this particular case implies
\begin{equation}
g_{ij}(\n)  = 
{1\over2} {\partial^2 [\bar q_2(\n,\n) \mp \sqrt{ \bar q_2(\n,\n)^2 - \bar q_0(\n,\n)\;(\n \cdot \n) }]\over \partial n^i \, \partial n^j}.
\end{equation}
It is convenient to rewrite the quadratics as
\begin{equation}
\bar q_2(\n,\n) = [\bar q_2]_{ij}\; n^i\,n^j; 
\end{equation}
\begin{equation}
\bar q_0(\n,\n) = [\bar q_0]_{ij}\; n^i\,n^j;
\end{equation}
since then we see
\begin{eqnarray}
g_{ij}(\n) &=&  [\bar q_2]_{ij} 
\mp \hbox{(discriminant contributions)}.
\end{eqnarray}
Unfortunately we shall soon see that the contributions coming from
the discriminant are both messy, and in certain directions,
ill-defined. This is obvious from the fact that squares
of both Finsler functions are not even everywhere differentiable.

Similarly we can construct a Finsler co-metric:
\begin{equation}
h^{ij}(\k) =  {1\over2} {\partial^2 [G_{3\pm}(\k)^2]\over \partial k_i \, \partial k_j},
\end{equation}
which specializes to
\begin{equation}
h^{ij}(\k) = 
{1\over2} {\partial^2 [q_2(\k,\k) \mp \sqrt{q_2(\k,\k)^2 - \bar q_0(\k,\k)\;(\k \cdot \k) }]\over \partial k_i \, \partial k_j}.
\end{equation}
It is again convenient to rewrite the quadratics as
\begin{equation}
q_2(\k,\k) = [q_2]^{ij}\; k_i\, k_j; 
\end{equation}
\begin{equation}
q_0(\k,\k) = [q_0]^{ij}\; k_i\,k_j;
\end{equation}
since then we see
\begin{eqnarray}
h^{ij}(\k) &=&  [q_2]^{ij} 
\mp \hbox{(discriminant contributions)}.
\end{eqnarray}
Again we shall soon see that the contributions coming from the discriminant are, in certain directions, problematic.

\subsection{Technical problems with the Finsler 3-metric}

Consider the (ray) discriminant
\begin{equation}
\bar D = \bar q_2(\n,\n)^2 - \bar q_0(\n,\n)\;(\n \cdot \n).
\end{equation}
There are three cases of immediate (mathematical) interest:

{\bf Isotropic:} If $v_x=v_y=v_z$ then $\bar D=0$; in this case the two Finsler functions $F_\pm$ are equal to ech other. $F_3(\d\x)$ then describes an ordinary Riemannian geometry, and $F_4(\d X)$ an ordinary pseudo--Riemannian geometry. This is the standard situation, and is for our current purposes physically uninteresting.

{\bf Uni-axial:} If one of the principal velocities is distinct from the other two, then  we can without loss of generality set $v_x=v_y = v_o$ and $v_z=v_e$.  The discriminant then factorizes into a perfect square
\begin{equation}
\bar D = \left\{ {(v_o^2-v_e^2)(n_x^2+n_y^2)\over 2 v_o^2 v_e^2 } \right\}^2.
\end{equation}
In this case it is immediately clear that both $F_{3\pm}(\d\x)$ reduce to simple quadratics, and so describe two ordinary Riemannian geometries. Indeed
\begin{equation}
F_{3+}(\n) = {\n\cdot \n\over v_o^2}; 
\end{equation}
and
\begin{equation}
F_{3_-}(\n) = {n_x^2+n_y^2\over v_e^2} + {n_z^2\over v_o^2}.
\end{equation}
In the language of crystal optics $v_o$ and $v_e$ are the ``ordinary'' and ``extraordinary'' ray velocities of a uni-axial birefringent crystal.  In geometrical language  the two photon polarizations ``see'' distinct Riemannian 3-geometries $F_{3\pm}(\d\x)$ and distinct pseudo-Riemannian 4-geometries $F_{4\pm}(\d X)$ --- this situation is referred to as ``bi-metric''. This situation is for our current purposes physically uninteresting.

{\bf Bi-axial:} The full power of the Finsler approach is \emph{only} needed for the bi-axial situation where the three principal velocities are distinct. This is the \emph{only} situation of real physical interest for us, as it is the only situation that leads to a non-trivial Finsler metric. In this case we can without loss of generality orient the axes so that $v_x>v_y>v_z$.  There are now two distinct directions in the $x$--$z$ plane where the discriminant vanishes --- these are the called the (ray) optical axes. After some manipulations that we relegate to \ref{A:axes}, the discriminant can be factorized as [\emph{cf.} equation~(\ref{E:duval})] 
\begin{equation}
\label{E:discriminant}
\bar D =
{(v_x^2-v_z^2)^2\over 4 v_x^4 v_z^4}
\times
\left[  (\n\cdot \n) - (\bar\e_1\cdot\n)^2 \right]
\left[  (\n\cdot \n) - (\bar\e_2\cdot\n)^2 \right],
\end{equation}
where the two distinct  (ray) optical axes are
\begin{equation}
\bar \e_{1,2} =
\left( \pm { {v_y\over v_x} \sqrt{v_x^2-v_y^2\over v_x^2-v_z^2}}; \;\; 0 \;\; ;
 { {v_y\over v_z} \sqrt{v_y^2-v_z^2\over v_x^2-v_y^2}}  \right).
\end{equation}
Note that $\bar \e_{1,2}$ are unit vectors (in the ordinary Euclidean norm) so that the discriminant $\bar D$ vanishes for any $\n\propto \bar\e_{1,2}$, and does not vanish anywhere else.  We can thus  introduce projection operators $\bar P_1$ and $\bar P_2$ and write
\begin{equation}
\bar P_1(\n,\n) =  (\n\cdot \n) - (\bar\e_1\cdot\n)^2; 
\end{equation}
\begin{equation}
\bar P_2(\n,\n) =  (\n\cdot \n) - (\bar\e_2\cdot\n)^2;
\end{equation}
Combining this with our previous results:
\begin{eqnarray}
\left\{ F_{3\pm}(\n)\right\}^2 &=& \bar q_2(\n,\n) 
\mp {(v_x^2-v_z^2)\over 2 v_x^2 v_z^2}
\sqrt{
\bar P_1(\n,\n) \; \bar P_2(\n,\n)
}.\quad
\end{eqnarray}
If we now calculate the Finsler metric $[g_{3\pm}(\n)]_{ij}$ we shall rapidly encounter technical difficulties due to the discriminant term. 
To make this a little clearer, let us define
\begin{equation}
[\bar P_3(\n)]_{ij} =
{\partial^2 \sqrt{ \vbox to 10.5pt{\null}
\bar P_1(\n,\n) \; \bar P_2(\n,\n) } \over \partial n^i \; \partial n^j},
\end{equation}
since then
\begin{equation}
[g_3(\n)]_{ij}  = [\bar q_2(\n)]_{ij}
\mp   {(v_x^2-v_z^2)\over 2 v_x^2 v_z^2}  [\bar P_3(\n)]_{ij}.
\end{equation}
Temporarily suppressing the argument $\n$, we have
\begin{equation}
[\bar P_3]_{ij} =  {1\over2} \partial_i \left[ \; \partial_j  \bar P_1 \;\sqrt{ \bar P_2\over  \bar P_1}
+  \partial_j  \bar P_2 \;\sqrt{ \bar P_1\over  \bar P_2} \; \right].
\end{equation}
A brief computation now yields the rather formidable result
\begin{eqnarray}
[\bar P_3]_{ij} 
&=& {1\over2}  \left[ \; \partial_i \partial_j \bar P_1 \; \sqrt{\bar P_2\over \bar P_1}
+  \partial_i \partial_j \bar P_2\; \sqrt{\bar P_1\over \bar P_2} \; \right]
\nonumber
\\
&&
+ {1\over4} \left[  {\partial_i \bar P_1 \; \partial_j \bar P_2 + \partial_i \bar P_2 \; \partial_j \bar P_1 \over \sqrt{\bar P_1 \bar P_2}}
-
\partial_i \bar P_1 \; \partial_j \bar P_1 \; {\bar P_2^{1/2}\over \bar  P_1^{3/2}}
-
\partial_i \bar P_2\;  \partial_j \bar P_2 \; {\bar P_1^{1/2}\over \bar P_2^{3/2}}
\right]
\\
&=& {1\over2\sqrt{\bar P_1\bar P_2}}  \left[ \; \partial_i \partial_j \bar P_1 \; \bar P_2
+  \partial_i \partial_j \bar P_2\; \bar P_1 \; \right]
\nonumber
\\
&&
+ {1\over4\sqrt{\bar P_1\bar P_2}} \left[  \partial_i \bar P_1 \; \partial_j \bar P_2
+ \partial_i \bar P_2 \; \partial_j \bar P_1
-
\partial_i \bar P_1 \; \partial_j \bar P_1 \; {\bar P_2\over \bar P_1}
-
\partial_i \bar P_2\;  \partial_j \bar P_2 \; {\bar P_1\over \bar P_2}
\right].\qquad
\end{eqnarray}
From this expression it is clear that along either optical axis, (as long as the optical axes are distinct, which is automatic in any bi-axial situation), \emph{some} of the components of $[\bar P_3]_{ij}$, and therefore \emph{some} of the components of the Finsler metric $[g_{3\pm}]_{ij} = [\bar q_2]_{ij} \pm \hbox{(constant)} \times [\bar P_3]_{ij}$, will be \emph{infinite}.

To see this in an invariant way, let $\u$ and $\w$ be two 3-vectors and consider
\begin{equation}
[\bar P_3(\n)](\u,\w)  =  [\bar P_3]_{ij} \; u^i\; w^j.
\end{equation}
After a brief computation:
\begin{eqnarray}
[\bar P_3(\n)](\u,\w)  
&=& {1\over2\sqrt{\bar P_1(\n,\n)\,\bar P_2(\n,\n)}}
\left[ \; \bar P_1(\u,\w) \; \bar P_2(\n,\n)
+  \bar P_2(\u,\w)\; \bar P_1(\n,\n) \; \right]
\nonumber
\\
&&+ {1\over4\sqrt{\bar P_1(\n,\n)\,\bar P_2(\n,\n)}} \Bigg[  \bar P_1(\n,\u) \; \bar P_2(\n,\w)
+ \bar P_2(\n,\u) \; \bar P_1(\n,\w)
\nonumber
\\
&&
-
\bar P_1(\n,\u) \; \bar P_1(\n,\w)  \; {\bar P_2(\n,\n)\over \bar P_1(\n,\n)}
-
 \bar P_2(\n,\u)\;  \bar P_2(\n,\w)  \; {\bar P_1(\n,\n)\over \bar P_2(\n,\n)}
\Bigg].\qquad
\end{eqnarray}

\noindent
This quantity will tend to infinity as $\n$ tends to either optical axis provided:
\begin{itemize}

\item The optical axes are distinct. 

(If the optical axes are coincident then $\bar P_1=\bar P_2$ and so $\bar P_3$ degenerates to
\begin{equation}
[\bar P_3(\n)](\u,\w) \to \bar P_1(\u,\w) = \bar P_2(\u,\w).
\end{equation}
One recovers the [for our purposes physically uninteresting] result for a uni-axial crystal.)

\item One is not considering the special case $\u=\w=\n$. 

(In this particular special case $\bar P_3$ degenerates to
\begin{equation}
[\bar P_3(\n)](\n,\n) \to \sqrt{ \bar P_1(\n,\n)\; \bar P_2(\n,\n)},
\end{equation}
which is well-behaved on either optical axis.)
\end{itemize}

\noindent
In summary: 
\begin{itemize}
\item 
The spatial Finsler 3-metric is $[g_{3\pm}]_{ij}$ \emph{generically ill-behaved on either optical axis}. 
\item
This feature will also afflict the \emph{spacetime} pseudo-Finsler 4-metric $[g_{4\pm}]_{ab}$ defined by suitable derivatives of the Finsler 4-norm $F_{4\pm}$.
\item 
This particular feature is annoying, but seems only to be a technical problem to do with the specifics of crystal optics, it does not seem to us to be a critical obstruction the developing a space-time version of Finsler geometry. Ultimately it arises from the fact that the ``null conoid'' is given by a quartic; this leads to two topological cones that for topological reasons always intersect, this intersection defining the optical axes. 
\item It is the technical problems associated with the (3+1) \emph{spacetime} Finsler metric, to be discussed below, which much more deeply concern us.
\end{itemize}

\subsection{Technical problems with the co-Finsler 3-metric}

The phase discriminant
\begin{equation}
D = q_2(\k,\k)^2 - q_0(\k,\k)\;(\k \cdot \k),
\end{equation}
arising from the Fresnel equation (and considerations of the phase velocity) exhibits features similar to those arising for the ray discriminant. There are three cases: 

{\bf Isotropic:} If the crystal is isotropic, then $D=0$. (This again is for our purposes physically uninteresting.) 

{\bf Uni-axial:} If the crystal  is uni-axial, then $D$ is a perfect square
\begin{equation}
D = \left\{ {(v_o^2-v_e^2)(k_x^2+k_y^2)\over 2}   \right\}^2
\end{equation}
and so the co-Finsler structures $G_{3\pm}$ are both Riemannian:
\begin{equation}
G_{3+}(\k) = {v_o^2 \; \k\cdot \n}; 
\end{equation}
\begin{equation}
G_{3_-}(\k) = { v_e^2\,(k_x^2+k_y^2)} + {v_o^2\, k_z^2}.
\end{equation}
 (This situation again is for our purposes physically uninteresting.) 

{\bf Bi-axial:} Only in the bi-axial case are  the co-Finsler structures $G_{3\pm}$ truly Finslerian.
There are now two distinct (phase) optical axes (wave-normal optical axes) along which the discriminant is zero, these optical axes being given by
\begin{equation}
\hat\e_{1,2} =
\left( \pm \sqrt{v_x^2-v_y^2\over v_x^2-v_z^2}; \;\;0\;\; ; \sqrt {v_y^2-v_z^2\over v_x^2-v_z^2} \right),
\end{equation}
in terms of which the phase discriminant also factorizes
\begin{equation}
 D =  {(v_x^2-v_z^2)^2\over 4}
\left[ (\k\cdot\k) - \left(\k \cdot \e_1 \right)^2 \right]  \left[ (\k\cdot \k) - \left(\k \cdot \e_2 \right)^2 \right].
\end{equation}
The co-Finsler norm is then (now using projection operators $P_1$ and $P_2$ based on the phase optical axes $\e_{1,2}$)
\begin{eqnarray}
\left\{ G_{3\pm}(\k)\right\}^2 &=& q_2(\k,\k) \mp
{{(v_x^2-v_z^2)\over2}}
\sqrt{
P_1(\k,\k) \; P_2(\k,\k)
}.
\end{eqnarray}
The co-Finsler \emph{metric} is defined in the usual way
\begin{equation}
[h_{3\pm}]^{ij}(\k) = {1\over2} {\partial^2 [G_{3\pm}(\k)^2]\over\partial k^i \; \partial k^j}.
\end{equation}
This now has the interesting ``feature'' that some of its components are infinite when evaluated on the (phase) optical axes. That is: The co-Finsler 3-metric is $[h_{3\pm}]^{ij}$ generically ill-behaved on either optical axis. This feature will also afflict the pseudo-co-Finsler 4-metric $[h_{4\pm}]_{ab}$ defined by suitable derivatives of the Finsler 4-norm $G_{4\pm}$.

\subsection{Technical problems with the (3+1) spacetime interpretation}

The (3+1)-dimensional spacetime objects that are closest in spirit to Riemann's original suggestion,\cite{Riemann} see \ref{A:Riemann}, are the quantities
\begin{equation}
F_4(\d X)^4 = \left\{ F_{4+} (\d X) \; F_{4-} (\d X) \right\}^2;
\end{equation}
and
\begin{equation}
G_{4}(k)^4 =\left\{ G_{4+}(k) \; G_{4-}(k) \right\}^2.
\end{equation}
as defined in equation (\ref{E:4d-dc-F}) and (\ref{E:4d-dc-G}).
This is tantamount to taking
\begin{equation}
F_4(\d X) = \sqrt{  F_{4+} (\d X) \; F_{4-} (\d X) };
\end{equation}
and
\begin{equation}
G_{4}(k) =\sqrt{ G_{4+}(k) \; G_{4-}(k) }.
\end{equation}
Now $F_4(\d X)$ and $G_4(k)$ are by construction perfectly well behaved pseudo-Finsler and pseudo-co-Finsler \emph{norms}, with the correct homogeneity properties --- and with the nice and concise physical interpretation that the vanishing of $F_4(\d X)$ defines a double-sheeted
``signal cone'' that includes both polarizations, while the vanishing of $G_4(k)$ defines a double-sheeted ``dispersion relation''   (``mass shell'') that includes both polarizations. (Thus $F_4(\d X)$ and $G_4(k)$ successfully unify the ``on-shell'' behaviour of the signal cones in a Fresnel-like manner.) These objects are also the closest we can get to simply and cleanly implementing Riemann's $4^{th}$-root proposal in a Lorentzian-signature setting.

While this is not directly a ``problem" as such, the norms $F_4(\d X)$ and $G_4(k)$ do have the interesting ``feature" that they pick up non-trivial complex phases: Since $F_{4\pm}(\d X)^2$ is always real,  (positive inside the propagation cone, negative outside), it follows that  $F_{4\pm}(\d X)$ is either pure real or pure imaginary. But then, thanks to the additional square root in defining  $F_4(\d X)$, one has:
\begin{itemize}
\item  $F_4(\d X)$  is pure real inside both propagation cones.
\item  $F_4(\d X)$  is proportional to $\sqrt{i} = {(1+i)\over\sqrt2}$
between the two propagation cones.
\item  $F_4(\d X)$  is pure imaginary outside both propagation cones.
\end{itemize}
Similar comments apply to the pseudo-co-Finsler norm $G_4(k)$. 

A considerably more problematic point is this: In the usual Euclidean signature situation the Finsler norm is taken to be smooth everywhere except for the zero vector --- this is usually phrased mathematically as ``smooth on the slit tangent bundle''.  What we see here is that in a Lorentzian signature situation the pseudo-Finsler norm cannot be smooth as one crosses the propagation cones --- what was in Euclidean signature a feature that only arose at the zero vector of each tangent space has in Lorentzian signature grown to affect (and infect) all null vectors. The pseudo-Finsler norm is at best ``smooth on the tangent bundle excluding the null cones''. (In a mono-refringent case the squared norm, $F_4(\d X)^2$, is smooth across the propagation cones, but in the bi-refringent case one has to go to the fourth power of the norm, $F_4(\d X)^4$, to get a smooth function.)

{\bf A ``no go'' result:}
Unfortunately, when attempting to bootstrap these two reasonably well-behaved  \emph{norms} to pseudo-Finsler and pseudo-co-Finsler \emph{metrics} one encounters additional and more significant complications. We have already seen that there are problems with the spatial 3-metrics $[g_{3\pm}]_{ij}(\n)$ and $[h_{3\pm}]^{ij}(\k)$
on the optical axes, problems which are inherited by the single-polarization spacetime (3+1)-metrics $[g_{4\pm}]_{ab}(\n)$ and $[h_{4\pm}]^{ab}(\k)$, again on the optical axes.

But now the spacetime (3+1)-metrics
\begin{equation}
[g_4]_{ab}(n) =  {1\over2} {\partial^2 [F_4(n)^2]\over\partial n^ a \; \partial n^b},
\end{equation}
and
\begin{equation}
[h_4]^{ab}(k) = {1\over2} {\partial^2 [ G_4(k)^2]\over\partial k_a \; \partial k_b},
\end{equation}
both have (at least some) infinite components --- $[g_4]_{ab}(n)$ has infinities on the entire signal cone, and $[h_4]^{ab}(k)$ has infinities on the entire mass shell. Since the argument is essentially the same for both cases, let us perform a single calculation:
\begin{eqnarray}
g_{ab} &=& {1\over2} \; \partial_a \partial_b \sqrt{ [F_+^2 \; F_-^2 ]}
\\
&=& {1\over4} \partial_a \left[ \partial_b [F_+^2] \; {F_-\over F_+}
+  \partial_b [F_-^2]\;  {F_+\over F_-} \right]
\end{eqnarray}
so that
\begin{eqnarray}
g_{ab} 
&=& {1\over4}  \left[ \partial_a \partial_b [F_+^2] \; {F_-\over F_+} +
 \partial_b \partial_b [F_-^2]  \; {F_+\over  F_-} \right]
\nonumber
\\
&&+ {1\over2} \bigg[
{\partial_a F_+ \partial_b F_- + \partial_a F_- \partial_b F_+ }
-
\partial_a F_+ \partial_b  F_+ {F_-\over F_+}
-
\partial_a F_- \partial_b F_- {F_+\over F_-}
\bigg].
\end{eqnarray}
That is, tidying up:
\begin{eqnarray}
g_{ab} &=& {1\over2} \left[ (g_+)_{ab} {F_2\over F_1}
+  (g_-)_{ab} {F_1\over F_2} \right]
\nonumber\\
&&
+ {1\over2} \bigg[  {\partial_a F_+ \partial_b F_-+ \partial_a F_- \partial_b F_+}
-
\partial_a F_+ \partial_b F_+ {F_-\over F_+}
-
\partial_a F_- \partial_b F_- {F_+\over F_-}
\bigg].
\end{eqnarray}
The problem is that this ``unified''  metric $g_{ab}(n)$ has singularities on both of the signal cones.
The (relatively) good news is that the quantity $g_{ab}(n) \, n^a n^b = F^2(n)$, and so on either propagation cone $F\to 0$, so $F(n)$ itself has a well defined limit.
But now let $n^a$ be the vector the Finsler metric depends on, and let  $w^a$ be some \emph{other} vector.
Then
\begin{eqnarray}
 g_{ab}(n)  \, n^a w^b &=&  {1\over2} n^a w^b \partial_a \partial_b [F^2] 
 \\
 &=& {1\over2} w^b \partial_b [F^2] 
 \\
 &=&  {1\over2} w^b \partial_b \sqrt{F_+ F_-}
\\
&=& {1\over4} w^b \left[ \partial_b [F_+^2] \; {F_-\over F_+} +
\partial_b [F_-]^2 \; {F_+\over F_-} \right]
\\
&=& {1\over2} \bigg\{ (\,[g_+]_{ab} \, n^a w^b) \; {F_-\over F_+} 
+
 (\,[g_-]_{ab} \, n^a w^b]) \; {F_+\over F_-} \bigg\}.
\end{eqnarray}
The problem now is this: $g_+$ and $g_-$ have been carefully constructed to be individually well defined and finite (except at worst on the optical axes). But now as we go to propagation cone ``$+$'' we have
\begin{equation}
g_{ab}(n) \, n^a w^b \to {1\over2}  (\,[g_+]_{ab} \; n^a w^b) \;  {F_-\over 0}  = \infty,
\end{equation}
and as we go to the other propagation cone ``$-$'' we have
\begin{equation}
g_{ab}(n) \, n^a w^b \to {1\over2}  (\,[g_-]_{ab} \; n^a w^b) \;  {F_+\over 0}  = \infty.
\end{equation}
So at least some components of this ``unified'' pseudo-Finsler metric $g_{ab}(n)$ are unavoidably singular on the propagation cones. Related (singular) phenomena have previously been encountered in multi-component BECs, where multiple phonon modes can interact to produce Finslerian propagation cones.\cite{lnp}

Things are just as bad if we pick $u$ and $w$ to be \emph{two} vectors distinct from ``the direction we are looking in'', $n$. In that situation
\begin{eqnarray}
g_{ab}(n) \, u^a w^b &=& {1\over2} \Bigg[
g_+(u,w) \,{F_-\over F_+} + g_-(u,w) \,{F_+\over F_-}
\nonumber
\\
&&
+ {g_+(u,n) \, g_-(w,n) + g_+(w,n) \, g_-(u,n)\over F_+ \, F_-}
\nonumber
\\
&&
- g_+(u,n) \, g_+(w,n) \, {F_-\over F_+^3} 
- g_-(u,n) \, g_-(w,n) \, {F_+\over F_-^3}
\Bigg].
\end{eqnarray}
Again, despite the fact that both $g_+$ and $g_-$ have been very carefully set
up to be regular on the propagation cones (except for the known, isolated, and tractable 
problems on the optical axes), the ``unified'' metric $g_{ab}(n)$ is
unavoidably singular there --- unless, that is, you \emph{only}
choose to look in the $nn$ direction. 

If we give up the condition of ``intuitiveness'', so that we abandon Riemann's $4^{th}$-root proposal, then $F_4$ and $G_4$ do not just factorize into a product of powers of the Finsler functions for individual polarizations. We could then look for more complicated ways of building ``unified'' Finsler and co-Finsler structures (constrained mainly by giving the correct light propagation cones in the birefringent crystal), and we might be able to find an appropriate Finsler geometry (which might also fulfill some additional reasonable physical conditions). The necessary conditions for such a geometry are relatively easy to formulate, but because of lack of intuitiveness, and corresponding lack of direct physical motivation, the  physical meaning of such an approach is highly doubtful.\cite{thessalonika}

In short, our ``no go'' result is this: There is no natural and compelling way of assembling the individual Finsler metrics into a ``unified'' whole --- this despite the fact that it \emph{is} possible to assemble the individual Finsler norms into a ``unified'' whole that accurately reflects the behaviour expected from looking at the Fresnel equation or the on-shell dispersion relations.

\section{Discussion and Conclusions}

The results of this investigation have been rather mixed.  On the one hand we have seen how birefringent crystal optics is a good exemplar for providing a clean physical implementation of  the mathematical notion of a Finsler norm and Finsler metric. On the other hand we have also seen how even this rather straightforward physical model, with its very clear and direct physical interpretation, nevertheless leads to a number of technical mathematical difficulties that are of physical importance.
\begin{itemize}

\item While the 3-space Finsler and co-Finsler \emph{norms} corresponding to the two photon polarizations are always well-defined,  the  3-space Finsler and co-Finsler \emph{metrics} have undesirable properties on the optical axes --- ultimately this is due to the fact that the two photon polarizations have the same group velocity (and same phase velocity) only on the optical axes --- so that the structure of the indicatrix (the surface describing speed as a function of direction) is complicated (continuous but not differentiable) on the optical axes. This feature then also afflicts the 4-space pseudo-Finsler and pseudo-co-Finsler metrics that cam be constructed for each individual photon polarization. While annoying, this problematic feature in 3-space is at least physically isolated to the optical axes.

\item It is when one tries to ``unify'' the two polarization modes into a single structure that
deeper problems arise --- again the 4-space pseudo-Finsler and pseudo-co-Finsler \emph{norms} are certainly well defined (and are extremely close to Riemann's original conception of what a $4^{th}$-order geometry should look like), but now 
the 4-space pseudo-Finsler and pseudo-co-Finsler \emph{metrics}  are singular on the entire signal cone and mass shell respectively.  This problematic feature is intimately related to the fact that we are dealing with \emph{pseudo}-Finsler and \emph{pseudo}-co-Finsler geometries --- it is a ``divide by zero'' problem, associated with non-zero null vectors on the signal cone (and mass shell), that leads to singular values for metric components. The feature seems to us to be quite comprehensive and to be generic to any attempt at merging pseudo-Finsler metrics with multi-refringence.

\end{itemize}
We have worked through the mathematical aspects of the Finsler space interpretation of birefringent optics in some detail because this is a situation in which we \emph{know} and \emph{understand} the basic physics in considerable detail. Because of this, any problematic mathematical details or physics oddities we run across cannot be ascribed to unknown details of the physics --- any unusual features we encounter must be ascribed to our choice of mathematical formalism, and if they already arise in this very simple physical system, then any such unusual features are likely to be inherited by more complicated physical situations. (In particular, since our long range goal is to develop insight into ``quantum gravity phenomenology'', issues we encounter here are likely to re-occur there, possibly with even more virulence.)

Overall, the lessons learned form this particular model have led us to re-evaluate and re-assess the notion of what it means to have a Lorentzian-signature Finsler geometry, and in particular  what it means to have a Lorentzian-signature Finsler \emph{metric} --- and to re-assess the precise nature of the relationship between birefringence and bimetricity (or more generally multi-refringence and multi-metricity). 
While pseudo-Finsler spacetimes are certainly useful constructs,  in birefringent situations it does not appear possible to naturally and intuitively construct a  ``unified'' pseudo-Finsler spacetime such that the pseudo-Finsler metric is null on both ``signal cones'', but has no other zeros or singularities --- it seems physically more appropriate to think of physics as taking place in a single topological manifold that carries a number of distinct pseudo-Finsler metrics, one for each polarization mode.  

That is, the objects $F_{4\pm}(\d X)$ individually define perfectly good pseudo-Finsler \emph{norms} and \emph{metrics}, but the natural ``unified'' quantity $F_4(\d X)$, while it defines a perfectly good pseudo-Finsler \emph{norm}, leads to a seriously diseased pseudo-Finsler \emph{metric}. This is an important point, and we are still actively developing our ideas on this matter.

\section*{Acknowledgments}

This research was supported by the Marsden Fund administered by the
Royal Society of New Zealand.
JS was also supported by a Victoria University of Wellington postgraduate scholarship.

\appendix

\section{Bernhard Riemann and Finsler geometry}
\label{A:Riemann}
To set the stage, we point out that ``Finslerian'' concepts have rather long history. As early as 155 years ago, in Bernhard Riemann's 1854 inaugural lecture,\cite{Riemann} he made some brief speculations about possible extensions of what is now known as Riemannian geometry:
\begin{quote}
  {The next case in simplicity includes those manifolds in
    which the line-element may be expressed as the fourth root of a
    quartic differential expression.  The investigation of this more
    general kind would require no really different principles, but
    would take considerable time and throw little new light on the
    theory of space, especially as the results cannot be geometrically
    expressed\dots}
\end{quote}

\begin{quote}
  {\dots A method entirely similar may for this purpose be applied
    also to the manifolds in which the line-element has a less
    simple expression, \emph{e.g.}, the fourth root of a quartic
    differential.  In this case the line-element, generally speaking,
    is no longer reducible to the form of the square root of a sum of
    squares, and therefore the deviation from flatness in the squared
    line-element is an infinitesimal of the second order, while in
    those manifolds it was of the fourth order.}
\end{quote}
In more modern language, Riemann was speculating about distances being defined by  expressions of the form
\begin{equation}
\label{E:Riemann4}
\d s^4 = g_{abcd}  \; \d x^a \; \d x^b  \;\d x^c  \;\d x^d.
\end{equation}
Such manifolds, and their generalizations, have now come to be called
Finsler geometries.\cite{Finsler} In particular, Riemann's specific suggestion 
\begin{equation}
\label{E:Riemann4b}
\d s = \sqrt[4]{g_{abcd}  \; \d x^a \; \d x^b  \;\d x^c  \;\d x^d},
\end{equation}
is now known as a $4^{th}$-root distance, and naturally leads to the specific sub-class of Finsler norms known as $n^{th}$-root distances. These and more general Finsler spaces are well-known in the mathematics community,\cite{Finsler}\cdash\cite{Bejancu}
but considerably less common within the physics community. (See for example references \refcite{Asanov}--\refcite{Vacaru7}, \refcite{Siparov}--\refcite{Gibbons2} 
and \refcite{Duval}--\refcite{thessalonika}.)
Perhaps the most extensive use of pseudo-Finsler geometries has been within the ``analogue spacetime'' community\cite{analogue} where Finsler-like structures have arisen in the context of normal mode analyses,\cite{birefringence}$^,$\cite{normal1}\cdash\cite{normal2} in multi-component BEC acoustics,\cite{lnp}$^,$\cite{trieste}\cdash\cite{silke-phd} and in nematic liquid crystals.\cite{nematic}

\section{Basics of bi-axial bi-refringent optics}
\label{A:BW}
The basic optics reference we shall use is Born and  Wolf, \emph{Principles of Optics}.\cite{BW} In particular we shall focus on Chapter XV, ``Optics of crystals'', pages 790--818. See especially pages 796--798 and pages 808--811. Specific page, chapter, and section references below are to  the 7th (expanded) edition, 1999/2003.

Permeability $\mu$ is taken to be a scalar, permittivity $\epsilon_{ij}$ a $3\times3$ tensor. (This is an \emph{excellent} approximation for all known optically active media.) By going to the principal axes we can, without loss of generality, take $\epsilon_{ij}$ to be diagonal
\begin{equation}
\epsilon_{ij} = \left[ \begin{array}{ccc} \ep_x & 0 & 0 \\ 0 & \ep_y & 0 \\ 0 & 0& \ep_z \end{array} \right].
\end{equation}
We furthermore define ``principal velocities''
\begin{equation}
v_x = {c\over\sqrt{\mu \ep_x}}; \qquad v_y = {c\over\sqrt{\mu \ep_y}}; \qquad v_z = {c\over\sqrt{\mu \ep_z}}.
\end{equation}
Note (this is a tricky point that has the potential to cause confusion) that $v_x$ is \emph{not} the velocity of light in the $x$ direction --- since $\ep_x$ (and so $v_x$) is related to the properties of the electric field in the $x$ direction, the principal velocity $v_x$  is instead the velocity of a light wave whose electric field is pointing in the $x$ direction. That is, for light waves propagating in the $y$-$z$ plane, \emph{one} of the polarizations will propagate with speed $v_x$.

\subsection{Group velocity and ray equation}

The group velocity, $v_g$, in the framework used by Born and Wolf, is identical to the  ``ray velocity'', and is controlled by the so-called ``ray equation". See   (15.2.29), page 797. To set some conventions, $\hat\n$ will always denote a unit vector in physical space --- a unit with respect to the usual Euclidean norm, while $\n$ is a generic position in physical 3-space. In contrast, $\hat \k$ will be reserved for a unit wave-vector in the dual ``wave-vector space''.

Born and Wolf exhibit the ray equation in a form equivalent (Born and Wolf use $\t$ where we use $\n$) to:
\begin{equation}
{\hat n_x^2\over 1/v_g^2 - 1/v_x^2} +  {\hat n_y^2\over 1/v_g^2 - 1/v_y^2} +  {\hat n_z^2\over1/ v_g^2 - 1/v_z^2} = 0.
\end{equation}
Here the group velocity (ray velocity) is defined by looking at the energy flux and
\begin{equation}
\v_g = v_g \; \hat \n.
\end{equation}
We can rewrite this as
\begin{equation}
{\hat n_x^2 v_x^2\over v_g^2 - v_x^2} +  {\hat n_y^2 v_y^2 \over v_g^2 - v_y^2} +  {\hat n_z^2 v_z^2\over v_g^2 - v_z^2} = 0.
\end{equation}
This form of the ray equation encounters awkward ``division by zero'' problems when one looks along the principal axes, so it is advisable to eliminate the denominators by multiplying through by  the common factor $ (v_g^2 - v_x^2) ( v_g^2 - v_y^2) ( v_g^2 - v_z^2)$, thereby obtaining:
\begin{eqnarray}
&&
\hat n_x^2 v_x^2 ( v_g^2 - v_y^2) ( v_g^2 - v_z^2) +
\hat n_y^2 v_y^2 ( v_g^2 - v_z^2) ( v_g^2 - v_x^2)
\qquad
\nonumber\\
&& 
\quad 
+
\hat n_z^2 v_z^2 ( v_g^2 - v_x^2) ( v_g^2 - v_y^2) = 0.
\qquad
\end{eqnarray}
It is this form of the ray equation that, (because it is much better behaved), we shall use as our starting point.
Now this is clearly a quartic in $v_g$, and by regrouping it we can write
\begin{eqnarray}
&&v_g^4 \left[  \hat n_x^2 v_x^2  +  \hat n_y^2 v_y^2  + \hat n_x^2 v_z^2  \right]
\nonumber\\
&& -
v_g^2 \left[ \hat n_x^2 v_x^2 (v_y^2+v_z^2) +\hat n_y^2 v_y^2 (v_z^2+v_x^2) +\hat n_z^2 v_z^2 (v_x^2+v_y^2) \right]
\nonumber
\\
 && \qquad
+ \left[ v_x^2 v_y^2v_z^2 \right]
= 0.
\end{eqnarray}
Equivalently
\begin{eqnarray}
&&v_g^4 \left[
\hat n_x^2 v_y^{-2} v_z^{-2}  +
\hat n_y^2 v_z^{-2} v_x^{-2}  + \hat n_z^2 v_x^{-2} v_y^{-2}  \right]
\nonumber
\\
 &&
\qquad
- v_g^2 \big[
\hat n_x^2  (v_y^{-2}+v_z^{-2}) +\hat n_y^2 (v_z^{-2}+v_x^{-2}) 
\nonumber\\
&&
 \qquad\qquad 
+\hat n_z^2  (v_x^{-2}+v_y^{-2}) \big]
 +1
= 0.
\end{eqnarray}
Now define two quadratics (in terms of the three direction cosines $\hat n_i$)
\begin{equation}
\label{E:bar-q0}
\bar q_0(\hat\n,\hat\n) = \left[
\hat n_x^2 v_y^{-2} v_z^{-2}  +
\hat n_y^2 v_z^{-2} v_x^{-2}  + \hat n_z^2 v_x^{-2} v_y^{-2}  \right] ,
\end{equation}
\begin{eqnarray}
\label{E:bar-q2}
\bar q_2(\hat\n,\hat\n) &=& {1\over2}  \big[
\hat n_x^2  (v_y^{-2}+v_z^{-2}) +\hat n_y^2 (v_z^{-2}+v_x^{-2}) 
+\hat n_z^2  (v_x^{-2}+v_y^{-2}) \big] ,
\end{eqnarray}
then
\begin{equation}
v_g^2(\hat\n) = {\bar q_2(\hat\n,\hat\n) \pm \sqrt{ \bar q_2(\hat\n,\hat\n)^2 - \bar q_0(\hat\n,\hat\n)}\over \bar q_0(\hat\n,\hat\n)}.
\end{equation}
But, since $\hat\n$ is a unit vector, we could equally well rewrite this as
\begin{equation}
v_g^2(\hat\n) = {\bar q_2(\hat\n,\hat\n) \pm \sqrt{ \bar q_2(\hat\n,\hat\n)^2 - \bar q_0(\hat\n,\hat\n) \; (\hat\n \cdot \hat\n) }\over \bar q_0(\hat\n,\hat\n)}.
\end{equation}
In this form both numerator and denominator are manifestly homogeneous and quadratic in the components of $\hat\n$, so for any 3-vector $\n$ (now \emph{not} necessarily of unit norm) we can take the further step of writing
\begin{equation}
v_g^2(\n) = {\bar q_2(\n,\n) \pm \sqrt{ \bar q_2(\n,\n)^2 - \bar q_0(\n,\n)\;(\n \cdot \n) }\over \bar q_0(\n,\n)}.
\end{equation}
The function $v_g(\n)$ so defined is homogeneous of degree zero in the components of $\n$:
\begin{equation}
v_g( \kappa\, \n ) = v_g(\n) = v_g(\hat\n).
\end{equation}
The homogeneous degree zero property should remind one of the relevant feature exhibited by the Finsler metric. It is also useful to note that
\begin{equation}
{1\over v_g(\n)^2} =  {\bar q_2(\n,\n) \mp \sqrt{ \bar q_2(\n,\n)^2 - \bar q_0(\n,\n)\;(\n\cdot \n) }\over (\n\cdot\n)}.
\end{equation}

\subsection{Phase velocity and Fresnel equation}

In contrast, the phase velocity, in the framework used by Born and Wolf, is controlled by the so-called ``equation of wave normals",   also known as the ``Fresnel equation''. See equation (15.2.24), page 796. The relevant computations are similar to, but not quite identical to,  those for the group velocity.

Let us consider a plane wave $\exp(i[\k \cdot \x - \omega t])$ and define the phase velocity by
\begin{equation}
\v_p = v_p \; \hat \k = {\omega\over k} \; \hat \k
\end{equation}
then the Fresnel equation is equivalent (Born and Wolf use $\s$ where we use $\hat\k$)  to
\begin{equation}
{\hat k_x^2\over v_p^2 - v_x^2} +  {\hat k_y^2\over v_p^2 - v_y^2} +  {\hat k_z^2\over v_p^2 - v_z^2} = 0.
\end{equation}
This form of the equation exhibits ``division by zero'' issues if you try to look along the principal axes, so it is for many purposes better to multiply through by the common factor $ ( v_p^2 - v_x^2) ( v_p^2 - v_y^2) ( v_p^2 - v_z^2)$ thereby obtaining the equivalent of their equation 
(15.3.1) on page 806:
\begin{eqnarray}
&&
\hat k_x^2 ( v_p^2 - v_y^2) ( v_p^2 - v_z^2) +
\hat k_y^2 ( v_p^2 - v_z^2) ( v_p^2 - v_x^2) 
\nonumber\\
&&+
\hat k_z^2 ( v_p^2 - v_x^2) ( v_p^2 - v_y^2) = 0.
\end{eqnarray}
This is clearly a quartic in $v_p$ and by regrouping it, and using $\hat\k \cdot \hat\k=1$, we can write
\begin{eqnarray}
&&
v_p^4 -
v_p^2 \left[ \hat k_x^2 (v_y^2+v_z^2) +\hat k_y^2 (v_z^2+v_x^2) +\hat k_z^2 (v_x^2+v_y^2) \right]
\nonumber\\
&&
+ \left[ \hat k_x^2 v_y^2v_z^2 +\hat k_y^2 v_z^2v_x^2 +\hat k_z^2 v_x^2v_y^2 \right]
= 0.
\end{eqnarray}
Let us now define two quadratics (in terms of the direction cosines $\hat k_i$)
\begin{equation}
\label{E:q2}
q_2(\hat\k,\hat\k) = {1\over2} \left[ \hat k_x^2 (v_y^2+v_z^2) +\hat k_y^2 (v_z^2+v_x^2) +\hat k_z^2 (v_x^2+v_y^2) \right],
\end{equation}
and
\begin{equation}
\label{E:q0}
q_0(\hat\k,\hat\k) =  \left[ \hat k_x^2 v_y^2v_z^2 +\hat k_y^2 v_z^2v_x^2 +\hat k_z^2 v_x^2v_y^2 \right],
\end{equation}
so as a function of direction the phase velocity is
\begin{equation}
v_p^2(\hat\k) = q_2(\hat\k,\hat\k) \pm\sqrt{ q_2(\hat\k,\hat\k)^2 - q_0(\hat\k,\hat\k) }.
\end{equation}
This is very similar to the equations obtained for the ray velocity. In fact, we can naturally extend this formula to arbitrary wave-vector $\k$ by writing
\begin{equation}
v_p^2(\k) = {q_2(\k,\k) \pm\sqrt{ q_2(\k,\k)^2 - q_0(\k,\k) \; (\k\cdot \k) }\over (\k\cdot\k)}.
\end{equation}
This expression is now homogeneous of order zero in $\k$, so that
 \begin{equation}
v_p(\kappa  \, \k) = v_p(\k) = v_p(\hat \k).
\end{equation}
Again, we begin to see a hint of Finsler structure emerging.

\subsection{Connecting the Finsler and co-Finsler structures}

Connecting the ray-vector $\hat \n$ and the wave-vector $\hat \k$ in birefringent optics is rather tricky --- for instance, Born and Wolf provide a rather turgid discussion on page 798 --- see section 15.2.2, equations (34)--(39).  The key result is
\begin{equation}
{v_g(\n)\; \hat n_i\over v_g(\n)^2-v_i^2} = {v_p(\k) \; \hat k_i\over v_p(\k)^2-v_i^2},
\end{equation}
which ultimately can be manipulated to calculate $\hat \n$ as a rather complicated ``explicit'' function of $\hat \k$ --- albeit an expression that is so complicated that even Born and Wolf do not explicitly write it down. Unfortunately all the extra technical machinery provided by Finsler notions of norm and distance do not serve to simplify the situation. (The fact that phase and group velocities can be used to define quite distinct, and in some situations completely unrelated,  effective metrics has also been noted in the context of acoustics.\cite{trieste}\cdash\cite{bled})

\section{Optical axes}
\label{A:axes}
To find the ray optical axes we (without loss of generality) take $v_z>v_y>v_z$, and define quantities $\bar \Delta_\pm$ (this is of course the result of considerable hindsight) by:
\begin{equation}
v_x^2 = {v_y^2 \over 1- v_y^2  \; \bar \Delta_+^2}; \qquad
v_z^2 = {v_y^2 \over 1 + v_y^2 \; \bar \Delta_-^2}.
\end{equation}
Furthermore eliminate $\hat n_y$ by using
\begin{equation}
\hat n_y^2 = 1 - \hat  n_x^2 -\hat  n_z^2,
\end{equation}
then
\begin{eqnarray}
 \bar D &=& \bar q_2^2 - \bar q_0
\nonumber\\
&=&
 {1\over4}
\left[ (\hat n_x \bar\Delta_+ +\hat n_z \bar\Delta_-)^2 - (\bar\Delta_+^2 + \bar\Delta_-^2 ) \right]
\nonumber\\
&&
 \times
\left[ (\hat n_x \bar\Delta_+ -\hat n_z \bar\Delta_-)^2 - (\bar\Delta_+^2 + \bar\Delta_-^2 ) \right]\!\!.
\end{eqnarray}
Thus the (ray) optical axes are defined by
\begin{equation}
(\hat n_x \bar\Delta_+ \pm \hat n_z \;\bar\Delta_-)^2 = (\bar\Delta_+^2 + \bar\Delta_-^2 ).
\end{equation}
But thanks to the Cauchy--Schwartz inequality
\begin{equation}
(\hat n_x \bar\Delta_+ \pm \hat n_z \bar\Delta_-)^2 \leq (\hat n_x^2 + \hat n_z^2)  (\bar\Delta_+^2 + \bar\Delta_-^2 ) \leq (\bar\Delta_+^2 + \bar\Delta_-^2 ).
\end{equation}
Therefore on the (ray) optical axis we must have $\hat n_y=0$, and $(\hat n_x^2 + \hat n_z^2)=1$. So (up to irrelevant overall signs)
\begin{equation}
\bar \e_{1,2} =
\left( \pm { \bar\Delta_+ \over\sqrt{\bar\Delta_+^2 + \bar\Delta_-^2 }}; \;\; 0 \;\; ;
 { \bar\Delta_- \over\sqrt{\bar\Delta_+^2 + \bar\Delta_-^2 }}  \right),
\end{equation}
which we can recast in terms of the principal velocities as
\begin{equation}
\bar \e_{1,2} =
\left( \pm { \sqrt{1/v_y^2-1/v_x^2\over1/v_z^2-1/v_x^2}}; \;\; 0 \;\; ;
 { \sqrt{1/v_z^2-1/v_y^2\over1/v_z^2-1/v_x^2}}  \right),
\end{equation}
or
\begin{equation}
\bar \e_{1,2} =
\left( \pm { {v_z\over v_y} \sqrt{v_x^2-v_y^2\over v_x^2-v_z^2}}; \;\; 0 \;\; ;
 { {v_x\over v_y} \sqrt{v_y^2-v_z^2\over v_x^2-v_z^2}}  \right).
\end{equation}
These are the two ray optical axes. (Compare with equation (15.3.21) on p.~811 of Born and Wolf.)

A similar computation can be carried through for the phase optical axes.
We again take $v_z>v_y>v_z$, and now define
\begin{equation}
v_x^2 = v_y^2 + \Delta_+^2; \qquad  v_z^2 = v_y^2 - \Delta_-^2.
\end{equation}
Eliminate $\hat k_y$ by using
\begin{equation}
\hat k_y^2 = 1 - \hat  k_x^2 -\hat  k_z^2.
\end{equation}
Then
\begin{eqnarray}
D &=& q_2^2-q_4 
\\
&=& {1\over4}
\left[ (\hat k_x \Delta_+ + \hat k_z \Delta_-)^2 - (\Delta_+^2+\Delta_-^2) \right] 
\nonumber
\\
&&
\times 
\left[ (\hat k_x \Delta_+ - \hat k_z \Delta_-)^2 - (\Delta_+^2+\Delta_-^2) \right]\!\!.
\end{eqnarray}
This tells us that the discriminant factorizes, \emph{always}.
The discriminant vanishes if
\begin{equation}
(\hat k_x \Delta_+ \pm \hat k_z \Delta_-)^2 = \Delta_+^2+\Delta_-^2.
\end{equation}
But by the Cauchy--Schwartz inequality
\begin{equation}
(\hat k_x \Delta_+ \pm \hat k_z \Delta_-)^2
\leq (\hat k_x^2 + \hat k_z^2) ( \Delta_+^2+\Delta_-^2)
\leq  ( \Delta_+^2+\Delta_-^2).
\end{equation}
Thus on the (phase) optical axis we must have $\hat k_y=0$ and $ (\hat k_x^2 + \hat k_z^2) = 1$.
The two unique directions (up to irrelevant overall sign flips) that make the discriminant vanish are thus
\begin{equation}
\e_{1,2} = \left( \pm {\Delta_+\over\sqrt{ \Delta_+^2+\Delta_-^2}} ; \;\; 0 \;\; ;
{\Delta_-\over\sqrt{ \Delta_+^2+\Delta_-^2}}\right),
\end{equation}
which can be rewritten as
\begin{equation}
\e_{1,2} =
\left( \pm { \sqrt{v_x^2-v_y^2\over v_x^2-v_z^2}}; \;\; 0 \;\; ;
 { \sqrt{v_y^2-v_z^2\over v_x^2-v_z^2}}  \right).
\end{equation}
These are the two phase optical axes. (Compare with equation (15.3.11) on p.~810 of Born and Wolf.)



\end{document}